\documentstyle[12pt,aps,epsf]{revtex}
\input{epsf}
\begin{document}

\title{Fourier Synthesization of Optical Pulses and ``Polar'' Light}
\author{V. S. Zapasskii and E. B. Alexandrov}
\address{All-Russia Research Center ``Vavilov State Optical Institute'',
St. Petersburg, 199034 Russia}
\maketitle

\vskip40pt
\begin{abstract}
It is shown that the direct Fourier synthesization of light
beams allows one to create polarity-asymmetric waves, which
are able, in the process of nonlinear interaction with a
medium, to break its inversion symmetry.  As a result,
these "polar" waves may show the effect of optical rectification
in nonlinear centrosymmetric media by generating light-induced
dc electric polarization.  At the same time, the waves of this
type, due to their unusual symmetry properties, can be used for
detecting the direction and sign of a dc electric field applied
to the medium.  The prospects of application of polar waves to
data recording and processing are discussed.
\end{abstract}

\section{INTRODUCTION}

The problem of generation of ultrashort light pulses has drawn
considerable attention in recent years. This attention is a result
of unprecedented (subfemtosecond) temporal resolution and giant
intensities achieved with their use, a potential possibility to
significantly increase the data transmission and processing rates,
and the unexpected successes in application of comb spectra of
ultrashort light pulses in metrology of optical frequencies (see,
e.g., the review \cite{1}). Since almost all existing methods for
producing ultrashort pulses involve generation of equidistant pulse
trains, the generated emission is also described by a discrete set
of equidistant harmonics in the spectral representation. In
practice, this is achieved either due to proper spectral
characteristics of the emitting system, e.g., an equidistant
cavity-mode structure in laser mode locking \cite{2}, or due to a
natural temporal periodicity of the physical process, as, e.g., in
the well-known experiments on generation of comb spectra under
high-power photoexcitation of noble gas atoms \cite{3,4}.

To address the problem of shortening the light pulse duration, Hansch
proposed in 1990 an idea of a direct synthesization of a wave with
the necessary field profile from a corresponding set of Fourier
harmonics [5]. Indeed, a train of ultrashort pulses may be obtained
by superposing several harmonics ($\omega_0, 2\omega_0,
3\omega_0,... n\omega_0$), with the highest of them $(n\omega_0)$
controlling the pulse duration and the lowest ($\omega_0$)
controlling the repetition rate. The ``light'' pulses synthesized
in this way are close in shape to a half-cycle of the highest
harmonic. The phases of the light beams being combined are supposed
to be sufficiently well defined, with the phases of different
harmonics synchronized due to their common origin from the same
master oscillator or due to phase-locking of different oscillators
(as in \cite{5}).

Figure 1 shows as an illustration the field
profile of two hypothetical waves synthesized from 30 harmonics of
the $CO_2$ laser (wave b only from odd harmonics). The highest
harmonic corresponds to the wavelength $\sim 0.35\hskip2pt \mu$m. The pulses
thus obtained are shorter than 1 fs and are spaced in time by $\sim
35$ fs. Implementation of such waves is an interesting and difficult
(as far as we know, so far unsolved) technical problem. In this
paper, however, we want to call attention to simple and unusual
properties of the synthesized waves which may be easily revealed
even with a minimal number of harmonics and may be interesting both
from the cognitive and practical points of view.

\section{POLAR
LIGHT WAVES AND THEIR PROPERTIES}

The two waves shown in Fig. 1
differ qualitatively from each other. In one of them (Fig. 1a) all
the pulses are unipolar, while in the other (Fig. 1b), they are
bipolar. The latter wave, in spite of the unusual shape of the
oscillations, still holds the main symmetry of a conventional
monochromatic wave with respect to sign reversal of the field in
each half-cycle.In the first ("unipolar") wave, the polar symmetry
of the oscillations is broken.

It is evident that in the processes
of linear interaction between the light and a medium for which the
superposition principle is obeyed the "unipolarity" of the wave
cannot be revealed because this property is distinguished only by a
certain phase matching of the constituent harmonics. However, in
the processes of a nonlinear interaction, the wave of this type may
show unique properties which are absent in its constituent
harmonics.

First of all, to describe adequately the physical
asymmetry of the unipolar wave field, the polarization vector of
the wave should become polar. It means that to describe
polarization state of a linearly-polarized wave one would have to
specify not only the azimuth of the polarization plane but also the
direction of its ``polarity'' (say, ``up'' or ``down''). Two beams
with ``parallel'' and ``anti-parallel'' polarizations are two
physically distinguishable situations. In particular, such beams
will interfere differently with each other. Figure 2 shows, as an
example, a possible shape of the polar wave synthesized from six
harmonics (the wave analyzed in \cite{5}) and the profile of the
interference pattern for the two waves of this kind with parallel
and anti-parallel polarizations. As is seen from the figure, these
two profiles are essentially different and, in particular,
completely destructive interference is possible only for the
interfering beams with antiparallel polarizations. The interference
pattern of this kind makes it possible to examine visually the
peculiar profile of the synthesized wave. Note, however, that such
an observation should be implemented with an appropriate broad-band
detector (e.g., an IR viewer), capable of detecting all spectral
harmonics of the interfering waves.

The most important property of
the polar waves, in our opinion, is their ability to induce, in a
centrosymmetric nonlinear medium, static electric polarization or
dc electric current, i.e., to induce a perturbation with the
symmetry of {\it polar vector} (see, e.g., \cite{6}). It is easy to
show that a result of combining the nonlinear polarizability of a
centrosymmetric medium ($P = \alpha E + \beta E^3 + \dots$) with the
polar asymmetry of the wave $E$ is that the integral of the
polarization $P$ over a period becomes nonzero. For
noncentrosymmetric media (with quadratic nonlinearity), the effect
of this kind is known as optical rectification \cite{7}. In the case
under consideration, the inversion symmetry of the problem is
lifted by the light wave itself. For a polar wave, the direction of
the induced dipole vector will depend on the azimuth of the plane
of polarization of the wave and will be evidently inverted upon
rotation of the polarization plane through $180^o$. Such a linearly
polarized wave is capable of bringing into the medium an electric
dipole moment exactly like a circularly polarized wave is capable
of bringing into the medium a magnetic moment. Unlike other methods
of light-induced charge separation in centrosymmetric media (as,
e.g., in the photorefractive effect \cite{8}, where the polar vector
that lifts the inversion symmetry is the gradient of the optical
excitation density), the use of polar waves allows one, in essence,
to locally create, in the illuminated spot, an effective electric
field with the magnitude and direction controlled by the intensity
and polarization of the light beam and thus to polarize the medium
in this area.

It seems evident that the polarization induced with
polar light can be detected (read out) using the same kind of
light. In other words, properties of a polarized medium (again, in
the nonlinear regime of interaction) will be different for the
beams with opposite directions of the polarization vectors.
Therefore, the polar wave that probes the polarized medium is
capable of detecting the sigh of its polarization or the sign of
the electric field applied to the medium. It may be easily shown
also that a usual monochromatic light wave after passing through a
polarized medium (e.g., placed in an external electric field)
acquires a "polarity", with the sign of this polarity (i.e., the
type of phase matching of the harmonics) being determined by the
sign of polarization of the medium, which can thus be detected. We
will consider these issues in a more rigorous way elsewhere.

It
should be emphasized that to obtain such polar light there is no
need to combine a great number of harmonics and to design exotic
waves like those presented in Fig. 1. Figure 3 shows profiles of
the polar waves synthesized from two and three harmonics. As is
seen, even for two harmonics, the field amplitudes of opposite
polarities may differ by a factor of two. Therefore, it may be
expected that even the simplest combinations of two harmonics will
be fairly efficient for observation of the above effects. Moreover,
one precedent of observation of the effect of this kind, when the
polar wave was formed and revealed in the same medium, is known.
This is optical frequency doubling in a glass optical fiber [9],
where the action of two coherent harmonics of laser light polarized
the medium within the regions of the order of the phase-matching
length and, thus, allowed the normally forbidden second harmonic
generation.

\section{NUMERICAL ESTIMATES}

Synthesized polar light waves provide fundamentally new
possibilities for perturbing and probing nonlinear media. The
possibility to locally polarize the medium with a light beam and to
subsequently read-out this polarization is very attractive from the
viewpoint of practical application of polar light, e.g., for data
recording and processing. This approach, however, has a
disadvantage: the medium can distinguish polarity of the wave only
when the process of interaction is nonlinear. Note that this
relates not only to data recording (which is inevitably nonlinear),
but also to read-out. For this reason, it is evident that
application of polar light for light-induced polarization of the
medium or for diagnostics of such a polarization may become
efficient, either in the case of high cubic nonlinearity of the
medium, or in the case of sufficiently high sensitivity of the
measurements. Let us consider this point in more detail.

As for
media with high cubic nonlinearity, one may rely on further
progress in the area of synthesis of new nonlinear materials based
on polymers with conjugated bonds where significant successes have
been achieved in recent years (see, e.g., \cite{10,11}). As can be
easily shown, the strength of the electric field $E_0$
produced in the medium with cubic susceptibility $\chi^{(3)}$ by a coherent
superposition of two harmonics with approximately equal amplitudes
$E$ is given, to within an order of magnitude, by ${\it E}_0 \sim
\chi^{(3)}E^3$. Using the scale of cubic nonlinearities achieved
nowadays in polymers with conjugated bonds, $\chi{(3)} \sim
10^{-10}$ CGSE (see, e.g., \cite{12}), one can estimate that the
amplitudes of the light wave harmonics needed to create in the
medium an effective dc field of about 100 V/cm will lie in the
range of $10^3$ CGSE. This corresponds to the light power density
of about $10^9$ W/cm$^2$ or $\sim$ 10 W through an area of 1 $\mu$m$^2$. It
should be borne in mind, however, that we made this estimate for
the most unfavorable case of superposition of two harmonics, when
the achievable polarity of the wave (which may be specified by the
ratio of amplitudes of the field vector in two opposite directions)
is the smallest. With increasing number of harmonics and increasing
polar asymmetry of the wave, the above estimate will become much
more favorable. It is possible also that considerable improvement
may be obtained by using semiconductors and quantum-confined
heterostructures with giant nonlinear susceptibilities (see, e.g.,
\cite {13}). For these cases, however, the relevant estimates
require a more serious analysis of the mechanism of polarizability
of the medium in a bichromatic field.

As applied to the problem of
{\it detecting} the electric polarization of the medium with the
polar-light waves, the situation appears much more attractive. The
point is that the spectral composition of polar light which has
passed through a polarized (by an external electric field)
nonlinear mediumproves to be different for the cases of parallel
and antiparallel orientations of the light and medium polarization
vectors. For simplicity, we will again consider superposition of
only two harmonics. From the viewpoint of the measurement
technique, the most interesting point is that the {\it amplitudes}
of these two harmonics appear to be odd functions of the beam
polarity. In other, more exact, words, when the sign of polarity of
each of the beams is inverted, the amplitude increment of each
harmonic also changes its sign. Beam polarity, in turn, may be
easily modulated by transmitting the beam through a medium with a
modulated (e.g., by an applied field) refractive index. Due to
dispersion of the medium, the difference between its refractive
indices for the two frequencies spaced by one octave ($\omega$ and
$2\omega$) will also be modulated and, for a properly chosen
modulation amplitude and thickness of the medium, the beam polarity
will appear to be modulated as well. As a result, as noted above,
the amplitudes of the constituent harmonics of the light beam that
has passed through the polarized medium under study will be
modulated. It is important that in an experiment of this kind, when
the beam is modulated in such a delicate fashion, the sensitivity
of measurements may easily reach its shot-noise limit. As can be
shown for this case, relative modulation of the fundamental
harmonic, will be given by the factor $6\chi^{(3)}{\it E_0}
E/\chi^{(1)}$, where $\chi^{(1)}$ and $\chi^{(3)}$ are the linear and
cubic susceptibilities of the medium, respectively, ${\it E_0}$ is
the dc field applied to the medium, and E is the field of the light
wave. Now, to estimate the limiting sensitivity of such a
technique, we have to compare this value with photocurrent
fluctuations of the detector, which registers the same light with
the amplitude $E$. As a result, the equation that connects the
strength of the applied dc electric field with the field amplitude
of the polar wave capable of detecting the corresponding
polarization of the medium will have the form .
$$
6\chi^{(3)}{\it E_0}E/\chi^{(1)} \sim 2\cdot
10^{-11}/(E\cdot\sqrt{S})
$$
Here $S$ is the beam cross section in
the medium. If we again take cubic susceptibility $\chi^{(3)} \sim
10^{-10}$ CGSE and the dc field applied to the medium equal to 30
CGSE ($\sim$ 10 kV/cm), it appears that this polarization of the medium
can be detected, e.g., in a 3-mW light beam with a cross section of
1 mm$^2$ (the power density 0.3 W/cm$^2$) or in a 3-$\mu$W light
beam with a cross section of 1 $\mu$m$^2$ (300 W/cm$^2$). As
before, sensitivity may be further increased by increasing the
number of harmonics that form the polar light beam. It is
noteworthy that in the above experiment with detection of
polarization of the medium, modulation of the light beam polarity
can be replaced by modulation of the sign of the field applied to
the medium. Of course, the result will be the same. However, the
possibility of such an observation indicates that with polar Light,
a centrosymmetric medium could exhibit a {\it linear} electrooptic
effect.

\section{CONCLUSIONS}

In this paper, we call attention to the fact that polar waves which
may be easily synthesized from a few Fourier harmonics of a
fundamental frequency may produce, in a nonlinear medium, a
perturbation with the symmetry of a polar vector (as, e.g., the
electric field or current). The action of such light beams upon a
nonlinear medium is equivalent to local application of an electric
field to the illuminated spot. In our opinion, this effect is
primarily interesting from the point of view of symmetry, as a way
to control and determine the {\it sign} of the polar vector using a
probe light beam. The estimates made above show that the method of
diagnostics of polarization of the medium using polar light may be
fairly efficient even for a medium with a small cubic nonlinearity.
The authors gratefully acknowledge useful discussions with S.G.
Przhibel'skii.

\begin{figure}
\epsfxsize=400pt
\epsffile {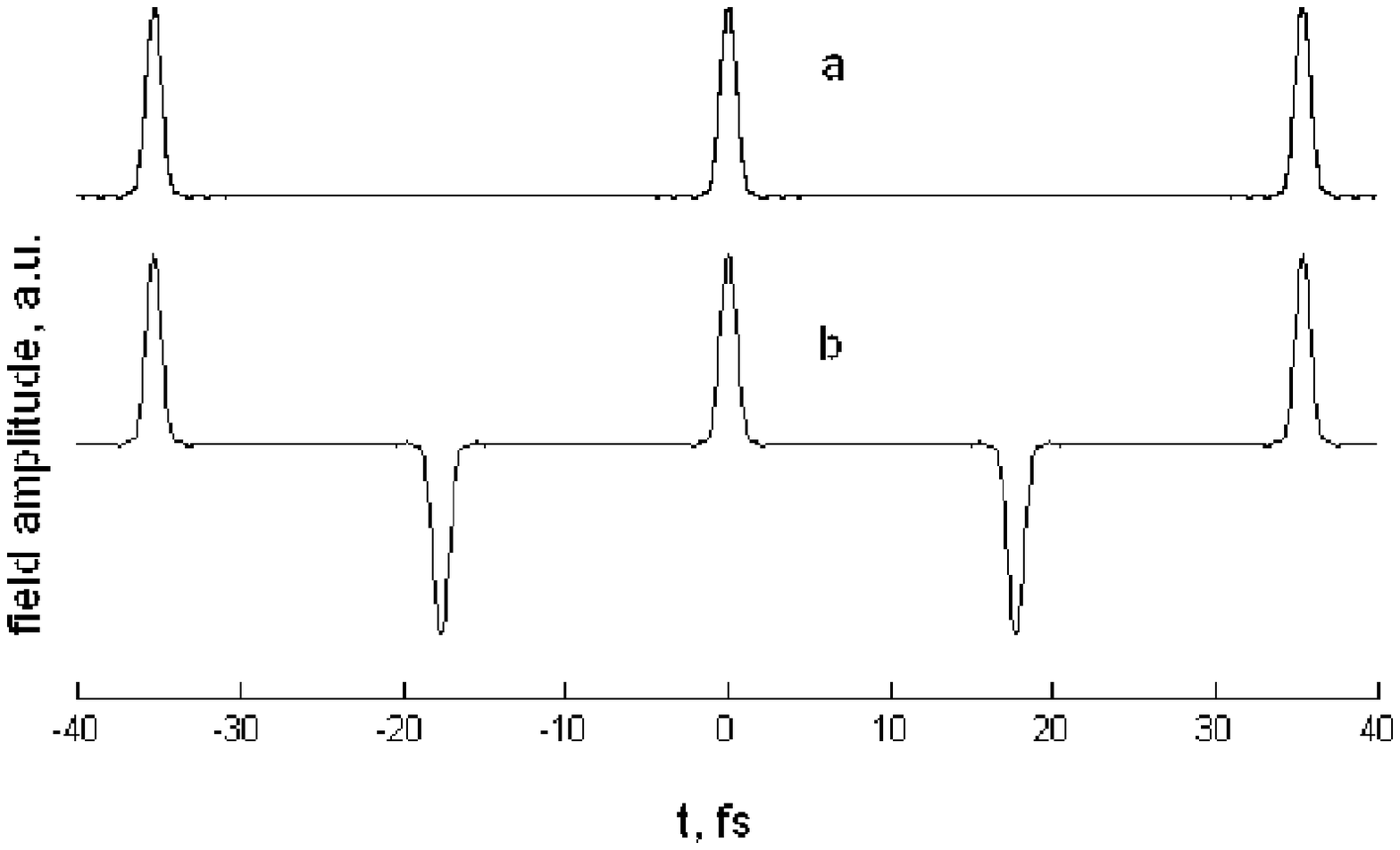}
\caption{The field profiles of two hypothetical waves synthesized from 30
harmonics of the $CO_2$ laser radiation ($\lambda = 10.6 \mu$m); a - unipolar wave and
b - bipolar wave.  The wave b is formed only from odd harmonics of the
fundamental frequency.}
\end{figure}

\begin{figure}
\epsfxsize=400pt
\epsffile {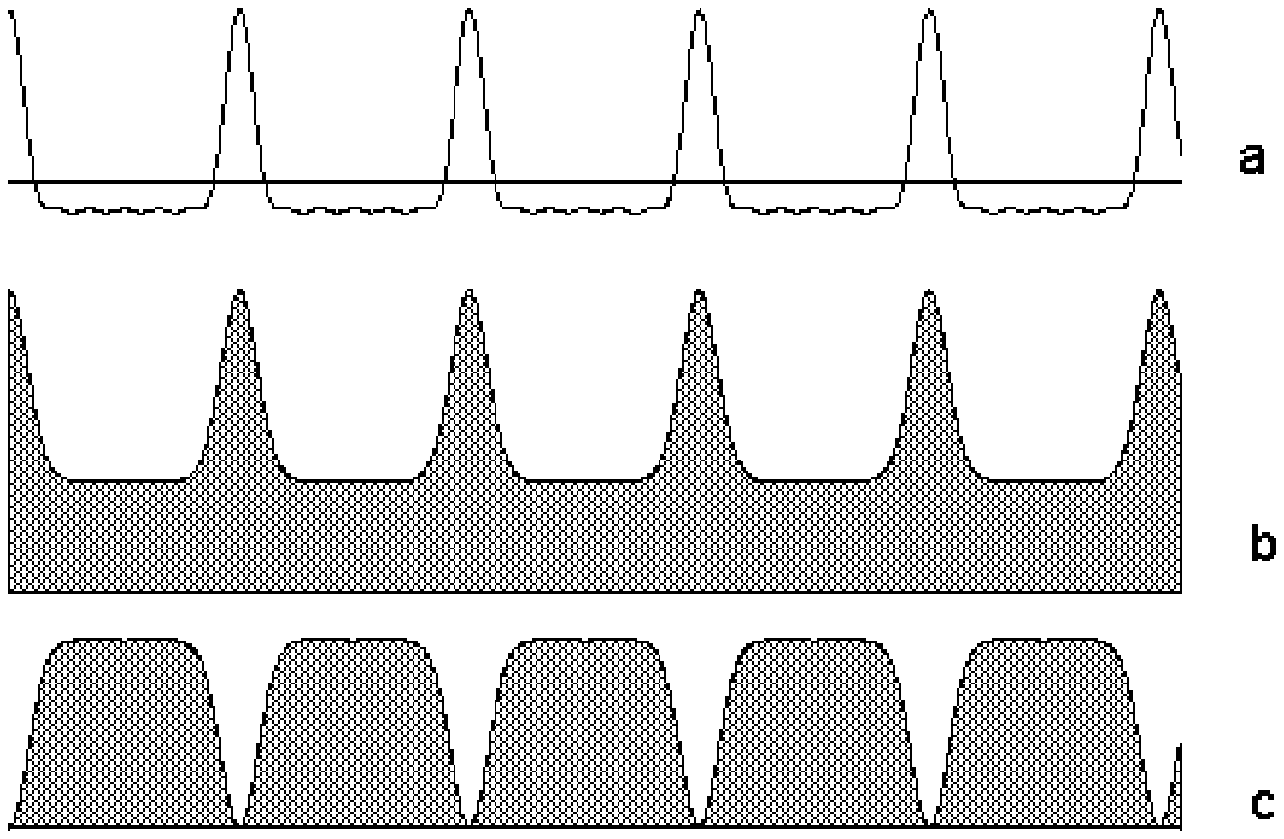}
\caption{ The field profile of the wave synthesized from six Fourier harmonics (a)
and the pattern of interference of two such waves with parallel (b) and
antiparallel (c) polarizations.}
\end{figure}

\begin{figure}
\epsfxsize=400pt
\epsffile {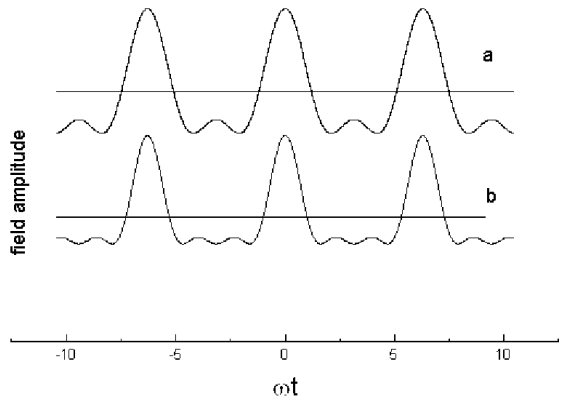}
\caption{ The field profile of the polar waves synthesized from
two (a) and three (b) harmonics.}
\end{figure}

\end{document}